\documentclass[12pt,prb,aps,dcolumn,superscriptaddress,showpacs,amsfonts,amsmath,amssymb,showkeys,floatfix]{revtex4-1}
\usepackage{bm}
\usepackage{pstricks}
\usepackage{setspace}
\usepackage{graphicx}
\begin{document}
\title{Evidence against an Almeida-Thouless line in disordered systems of Ising dipoles}
\date{\today}
\author{Julio F. Fern\'andez}
\affiliation{Instituto de Ciencia de Materiales de Arag\'on, CSIC-Universidad de Zaragoza, 50009-Zaragoza, Spain}
\affiliation{Instituto Carlos I de F\'{\i}sica Te\'orica y Computacional,  Universidad de Granada, 18071 Granada, Spain}

\date{\today}

\begin{abstract}
By tempered Monte Carlo simulations, an Almeida-Thouless (AT) phase-boundary line in site-diluted Ising spin systems
is searched for. Spins  interact only through dipolar fields and occupy
a small fraction of lattice sites. 
The spin-glass susceptibility of these systems and of the Sherrington-Kirkpatrick model are compared. 
The correlation length as a function of system size and temperature is also studied.
The results obtained are contrary to the existence of an AT line. 
\end{abstract}

\maketitle

\section{Introduction}
\label{intro}

The collective behavior of some spin systems is controlled by dipole-dipole interactions. It is so in some magnetic nanoparticle \cite{nanoscience} arrays,\cite{np, sachan}  in some crystals of organometallic molecules, \cite{nanomag}
as well as in some magnetic salts, such as LiHoF$_4$. In LiHoF$_4$, uniaxial
crystal-field anisotropy forces the Ho ion spins to point up or down along the anisotropy axis.\cite{HoY,grif} A model of Ising spins with dipole-dipole interactions
ought therefore to capture the main features of the magnetic behavior of LiHoF$_4$. This system  orders 
ferromagnetically at low temperatures, which, as  Luttinger and Tisza\cite{lutt} showed long ago, is accidental. Had the Ho ions crystallized in a simple cubic lattice, for instance, it would have ordered
antiferromagnetically.\cite{odip} 
This illustrates how delicate the balance between dipolar fields coming from different sources is. 
The frustration that underlies such a balance is expected to lead to spin-glass behavior in disordered-Ising-dipole (DID) models which mimic the LiHo$_x$Y$_{1-x}$F$_4$ family of materials\cite{bel} if $x\ll 1$. 

Some details about  LiHo$_x$Y$_{1-x}$F$_4$, such as the symmetry of its crystalline lattice, are irrelevant\cite{alonso2010} if $x\ll1$. Other details, such as transverse fields, which have no place in the DID model, do make a difference. Thus, interesting quantum effects that have been observed\cite{ultimoq,barbara} in LiHo$_x$Y$_{1-x}$F$_4$ at low temperatures are beyond DID models. On the other hand, a clear picture of the DID model seems like a good starting point for the study of quantum dipolar systems.  Thus far, no such clear picture exists. 

Several experiments\cite{ultimoq,rosen} on LiHo$_x$Y$_{1-x}$F$_4$ suggest there is a paramagnetic (PM) to spin glass (SG) phase transition when $x\lesssim 0.25$, but some skepticism remains.\cite{barbara} Some computer simultion of DID models\cite{yu} point to a PM phase for all nonvanishing temperatures. However, the opposite conclusion has been drawn more recently.\cite{gin,alonso2010}  

Below the transition, the nature of the hypothetical SG phase of DID models remains rather unexplored.  Simulations for zero applied field suggest\cite{alonso2010} the DID model behaves in three dimensions (3D) somewhat similarly to the XY model  in 2D. Thus, $3$ would be the value of the lower critical dimension $d_L$ of DID models in zero applied field. Note, however, (i) that the correlation length of the Edwards-Anderson (EA) model has previously been observed\cite{balle} to behave similarly, as a function of system size and temperature, (ii) that $d_L<3$ was nevertheless drawn from this behavior, and that (iii) this fits in with a $d_L\simeq 2.5$ value that has recently been inferred for the EA model from other evidence.\cite{stiff,earlydl,Fisch,kardar} 
I know of no reported work on the behavior of DID models under applied longitudinal magnetic fields.

\begin{figure}[!t]
\includegraphics*[width=80mm]{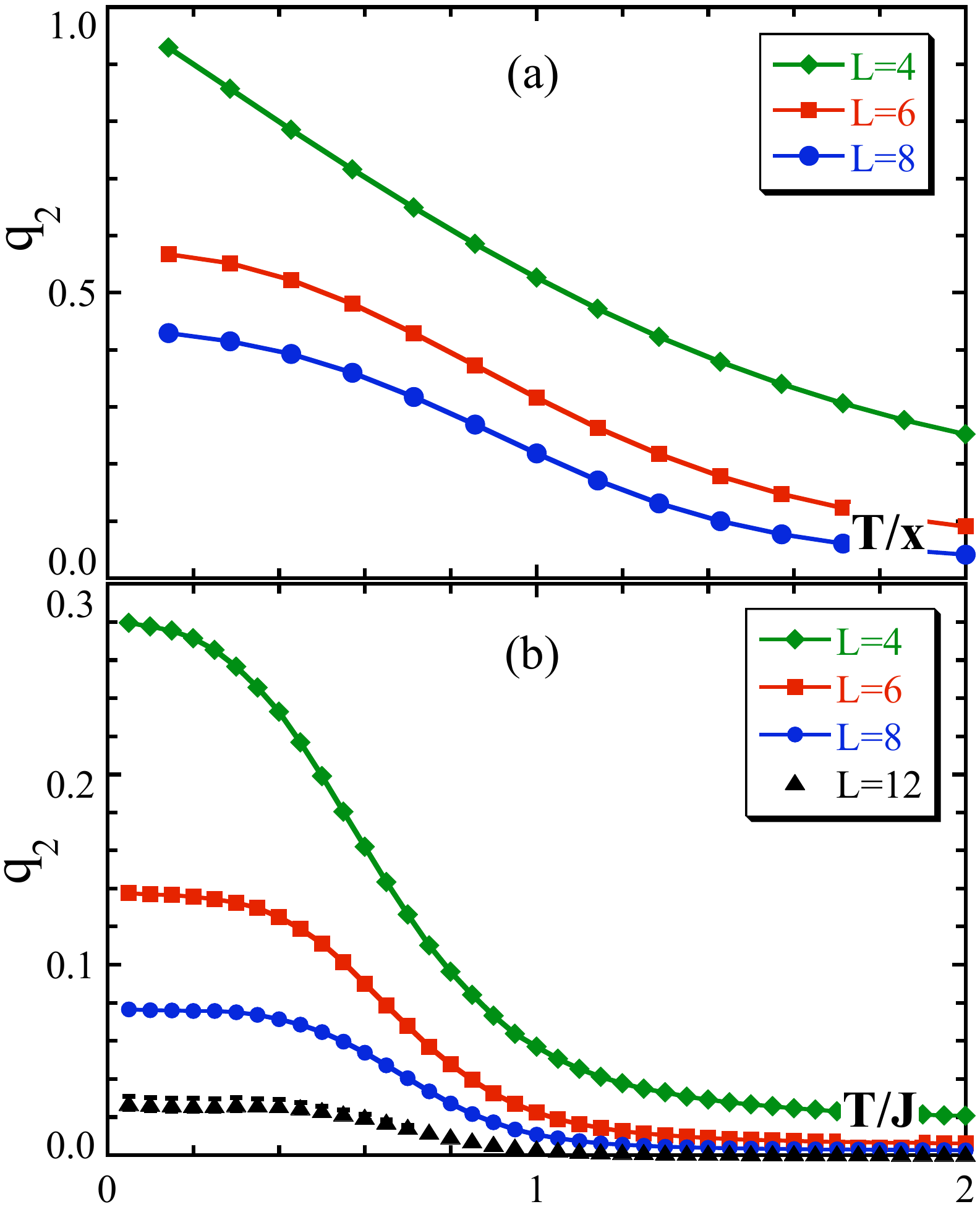}
\caption{(Color online) (a) Plots of $q_2$ vs $T/x$ for an $L^3$ DID model on a SC lattice, for $x=0.35$ and the $L$ values shown in the graph. Here, $q_2$ is for two replicas, both of which are in equilibrium but under different applied fields, $H_1=0$ and $H_2=0.2$. (b) Plots of $q_2$ vs $T/J$ for the $L^3$ SK model, the $L$ values shown in the graph, and $H_1=0$ and $H_2=0.4$ for replicas 1 and 2. In both (a) and (b), most error bars do not show because they hide behind icons. Lines are guides to the eye.}
\label{qv}
\end{figure}

Whether there is a thermal phase transition, between the PM and SG phases, as the temperature $T$ is lowered in an applied magnetic field $H$ is an important question. 
An $H-T$ phase-boundary line was long ago discovered in the Sherrington-Kirkpatrick\cite{sk} (SK) model by de Almeida and Thouless (AT).\cite{at} For its existence in the EA model, there is both favorable\cite{yesEA,yesmore,yes1d} and unfavorable\cite{noatea,jorgo,crm} evidence. To get a feeling for the physics involved, consider first the argument of 
Fisher and Huse,\cite{droplet2} which in turn follows from Imry and Ma's argument\cite{iandma} for the instability of diluted AFs to the application of a magnetic field. Consider a patch of $n$ spins in a SG state at $H=0$. Because all the nearest neighbor bonds are of random sign, the numbers of spins pointing in opposite directions are then expected to differ by $\sim n^{1/2}$. The Zeeman energy  therefore changes by $\Delta E_H\sim Hn^{1/2}$ if a patch of $n$ spins is flipped when $H\neq 0$. Let the corresponding energy change coming from broken bonds be given by $\Delta E_J\sim \Upsilon n^{\theta /3}$, which defines the stiffness\cite{stiff} constant $ \Upsilon $ and the stiffness exponent $\theta$. Fisher and Huse\cite{droplet2} further showed that $\theta \leq (d-1)/2$ for the EA model (more recent numerical work gives\cite{stiff2} $\theta\approx 1.2 \ln (0.4d)$ for $1\lesssim d <  6$), whence  $\Delta E_J<\Delta E_H$  follows for a sufficiently large value of $n$. 
Widespread spin reversals of this sort on macroscopic systems would lead to a state with a $q=0$ overlap with the initial state.   
(The spin overlap $q$ between two spin configurations may be defined as the total fraction of sites on which spins
point in the same direction minus the fraction of sites on which spins point oppositely.)  Because dipole-dipole interactions are long ranged, the above argument is not immediately applicable to the DID model. Data for the mean square $q_2$ of the overlap between equilibrium states at $H=0$ and at\cite{Hdef} $H=0.2$ is exhibited in Fig. \ref{qv}a for the DID model, for $x=0.35$, all $T$ and various system sizes in 3D. These results suggest that indeed $q\rightarrow 0 $ as $L\rightarrow \infty$ for the DID model as well. 
Analogous results are shown in Fig. \ref{qv}b for the SK model. Again,  $q\rightarrow 0 $ as $L\rightarrow \infty$ seems to ensue. This is in spite of the fact that  an AT line is known to exist for the SK model. 
Whereas Imry and Ma\cite{iandma} could conclude that a small magnetic field can destroy the antiferromagnetic phase of a dilute antiferromagnet (AF), the analogous conclusion could only be drawn for the DID model if it were known to fit the droplet scenario\cite{droplet2,droplet1} (in which there is no ground state degeneracy).  This is why Fig.  \ref{qv}a provides insufficient evidence for the nonexistence of the Almeida-Thouless line in the DID model.
An analogy with a simpler system is helpful at this point.

\begin{figure}[!t]
\includegraphics*[width=80mm]{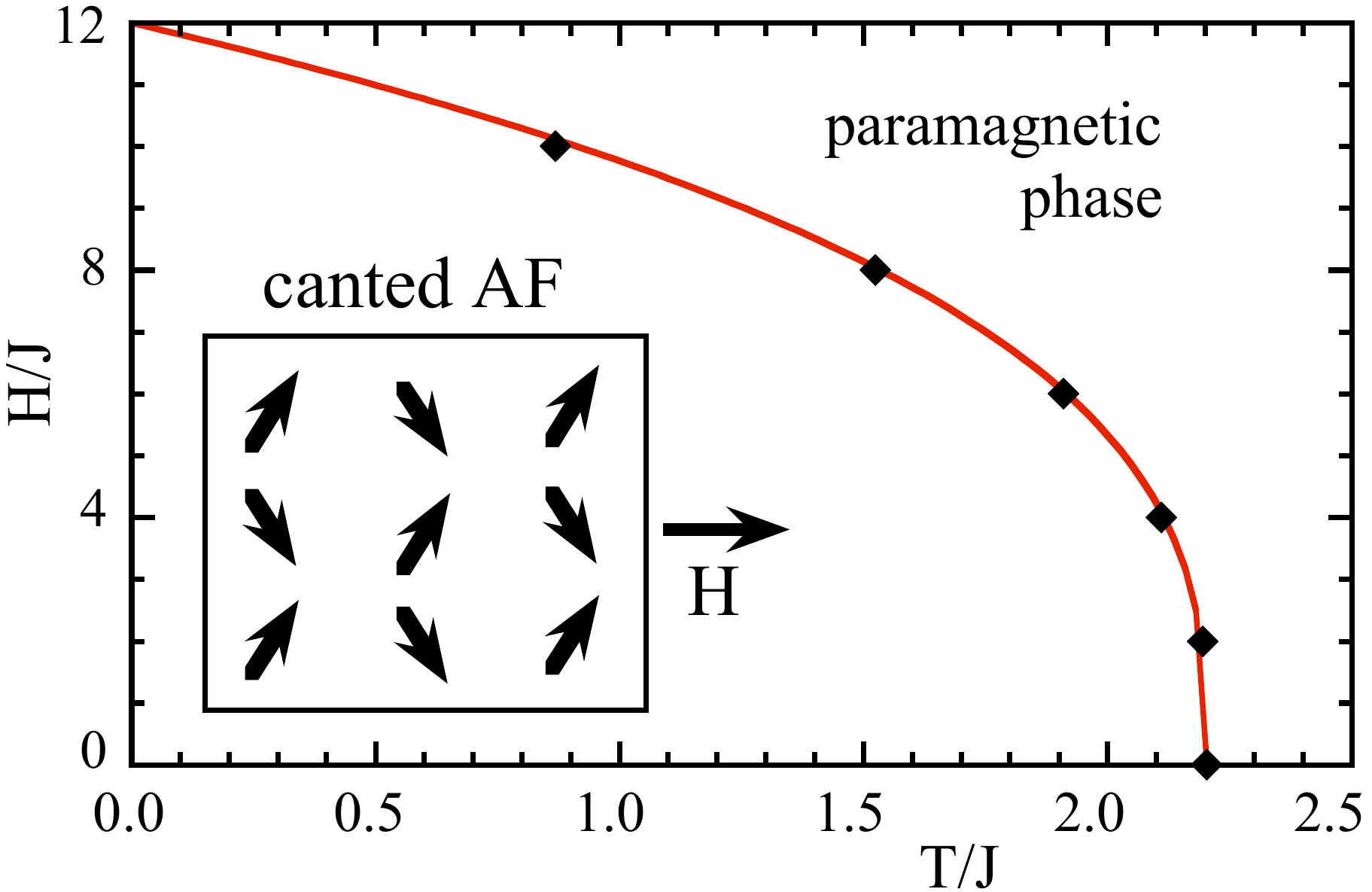}
\caption{(Color online) $H/J$ vs $T/J$, where $J$ is a nearest neighbor exchange constant, phase diagram of the antiferromagnetic XY model in 3D.  Data points come from MC simulations. The fitting curve, $H=11.8(1-T/T_c)^{0.37}$, where $T_c=2.20J$, 
 is also shown. }
\label{xy}
\end{figure}

Consider an isotropic AF. Upon the application of an arbitrarily small magnetic field $H$, all spins rotate uniformly till they point nearly perpendicularly to $H$. From a canted AF alignment, spins can better minimize the ground state energy. 
It takes a nonvanishing $H$ to further drive this ``spin-flop'' phase beyond the H-T boundary line, into the paramagnetic phase.\cite{neel} This is illustrated in Fig. \ref{xy} for the XY model in 3D.
The phase transition on the H-T boundary line can take place because the applied field does not completely lift the ground state degeneracy. Two degenerate states (for two sublattices) survive. 
Fluctuations between these two states enable the existence of an H-T boundary line.  Thus, \emph{sublattice symmetry} is broken below the H-T line. Analogously, critical fluctuations between various low energy states take place on an AT line. In the SG phase, different replicas of a SK system can stay on different states. 
This sort of replica equality breaking, is known as \emph{replica symmetry breaking}\cite{bray} (though no symmetry operation relates these states).

In the droplet scenario there are only two states, related by global spin inversion. An arbitrarily small magnetic field therefore lifts this degeneracy. Only one state survives, which leaves no room for critical fluctuations to occur at any nonzero $H$.  Hence,
Fisher and Huse\cite{droplet2,droplet1} concluded that $q =0$ between two states, one at $H=0$ and another one at $H\neq 0$, implies the state for $H\neq 0$ is not a SG state.  Thus, the nonexistence of an AT line is a clear cut prediction of the droplet model.

The aim of this paper is to establish whether there is an AT phase-boundary line in a site diluted DID model in 3D. 
This is to be done by means of the tempered Monte Carlo (MC) method.\cite{tempered} 
The behavior of the DID model, has been previously shown\cite{alonso2010} to depend on $x$ and $T$ only through $T/x$ for $x\ll 1$. 
It therefore suffices to study how the model behaves as a function of $T$ and $H$ at a single value of $x$. 

A brief outline of the paper follows. The DID model is defined in Sec. \ref{mo}. The boundary conditions are described in Sec. \ref{method}. The definition of the spin-overlap
parameter $q$ and how it is calculated can also be found in Sec. \ref{method}. How equilibration times of the 
DID model under tempered MC rules are arrived at is described in Sec. \ref{equil}. Equilibrium results for the spin-glass susceptibilities $\chi_{sg}$ of the DID model and SK models, both for $H=0$ and 
and $H\neq 0$, are compared in Sec. \ref{er}. Equilibrium results for the correlation length $\xi_L$ of the DID model are also given in Sec. \ref{er}. Results for both $\chi_{sg}$ and $\xi_L$ are clearly in accord with the absence of an AT phase-boundary line in the DID model. Further concluding remarks appear in Sec. \ref{concl}.

\section{model,  method and equilibration}
\label{mm}
\subsection{Model}
\label{mo}

The DID model on a simple cubic (SC) lattice is next defined. 
All dipoles point along the $z$ axis of the lattice. Each site is occupied with probability $x$. 
The Hamiltonian is given by,
\begin{equation}
{\cal H}=\frac{1}{2}\sum_{ ij}\sigma_i
T_{ij}\sigma_i -H\sum_i\sigma_i
\label{eq1}
\end{equation}
where the sums are over all occupied sites, except for $i=j$ in the double sum. 
$\sigma_i=\pm 1$ on all 
occupied sites $i$, 
\begin{equation}
T_{ij}=\varepsilon_a
(a/r_{ij})^3(1-3
z_{ij}^2/r_{ij}^2),
\label{T}
\end{equation} 
$ r_{ij}$ is the distance between $i$ and $j$ sites, $z_{ij}$ is the $z$ component of $ r_{ij}$, $\varepsilon_a$ is an energy, and $a$ is the
SC lattice constant. 

For $H=0$, the DID model has been shown\cite{alonso2010} to have an equilibrium PM-SG transition if $x< 0.65(5)$ (in SC lattices). Furthermore, the PM-SG transition temperature is given by $T_{sg}=1.0(1)x$ for all $x\lesssim 0.5$.

For comparison,  a few results for the SK model are also shown. Then, all exchange constants are given random values chosen independently  from the same Gaussian distribution, centered on $0$ with a $J^2/N$ mean square deviation.

Unless otherwise stated, all  temperatures and energies for the DID model are given in terms of $\varepsilon_a/k_B$ and $\varepsilon_a$, respectively. The magnetic field $H$ is defined by Eq. (\ref{eq1}) to be an energy, and is therefore also given in terms of $\varepsilon_a$. All times are given in MC sweeps (MCS).

\subsection{Method}
\label{method}
 
I use periodic boundary conditions (PBC), in which a periodic arrangement of replicas span all space beyond the system of interest.
These replicas are exact copies of the Hamiltonian and of the spin configuration of the system of interest.  
A spin on site $i$ interacts through dipolar fields with all spins within an $L\times L\times L$
cube centered on it. No interactions with spins beyond this cube are taken into account. (Additional details of the PBC scheme used here can be found in Ref. \onlinecite{odip}.)
This may seem odd, because dipolar interactions make themselves felt over macroscopic distances.
That is why different ``demagnetization factors'' apply to differently shaped macroscopic bodies.\cite{kittel}  On the other hand, demagnetization factors vary with system shape, but not with macroscopic system size. Indeed, the error that is introduced by this method was shown in Ref. \onlinecite{alonso2010} to vanish as $L\rightarrow \infty$, regardless
of whether the system is in the paramagnetic, AF or SG phase (but not near a ferromagnetic phase transition). This enables us to disregard interactions of any one spin on site $i$ with any spin beyond an $L\times L\times L$ cubic box centered on site $i$.

In order to bypass energy barriers that can trap a system's state at low temperatures
the parallel tempered MC algorithm is used here,\cite{tempered} following the steps outlined in Ref. \onlinecite{alonso2010}. 
Configuration swap rates between systems at temperatures $T$ and $T+\Delta T$ were checked to be reasonably large throughout.
The smallest swap rates ensue for the lowest temperature (i.e., $T=0.05$) and the largest systems (i.e., $L=10$). Then, swap rates in equilibrium were found to be approximately $0.3$, i.e., $30 \% $ of all attempts made for configuration exchanges are successful. Swap rates increase slowly with increasing $T$ in the spin-glass phase, and faster above $T_{sg}$.

In order to be able to calculate spin overlaps between different equilibrium states at the same temperature,
not one, but two sets, each one of $n$ identical systems, are allowed to evolve independently in parallel. All $2n$ systems start from independently chosen random configurations.
The temperature spacing $\Delta T$ between systems in each set was chosen to be $\Delta T=0.05$. Checks for equilibrium are described below, following the time dependent spin-overlap definitions.

As usual, the Edwards-Anderson overlap\cite{ea} between identical systems (replicas) $1$ and $2$ is defined by,
\begin{equation} 
q=N^{-1} \sum_j \phi_j,
\label{q}
\end{equation}
where 
\begin{equation}
\phi_j=\sigma^{(1)}_j\sigma^{(2)}_j,
\end{equation}
$\sigma^{1}_j$ and $\sigma^{2}_j$ are the spins on site $j$ of identical replicas $(1)$ and $(2)$ of the system of interest. Unless otherwise stated, identical replicas have, as usual, the same Hamiltonian. Exceptionally, for Figs. \ref{qv}a and \ref{qv}b, different fields $H_1$ and $H_2$ are assumed to be applied to replicas $1$ and $2$, respectively.

\subsection{ Equilibration}
\label{equil}

The purpose of this subsection is to establish how long it takes the DID model to come to thermal equilibrium. 
In order to be able to follow the equilibration process (under tempered MC rules), some useful quantities are next defined. 
First, two replicas are allowed to evolve independently, starting at $t=0$ from two uncorrelated \emph{random} states $\mathfrak{r}_\mu$ and  $\mathfrak{r}_\nu$. Let ${q}_2(t\mid \mathfrak{r}_\mu,\mathfrak{r}_\nu)$ be the average of $q^2$ at time $t$ over all sample realizations. Different samples start from different random pairs of states, $ \mathfrak{r}_\mu$ and $\mathfrak{r}_\nu$. 
In ${q}_2(t\mid \mathfrak{r}_\mu,\mathfrak{r}_\nu)$, $ \mathfrak{r}_\mu$ and $\mathfrak{r}_\nu$ appear only to remind us that all initial pairs of states at $t=0$ are uncorrelated random states.

\begin{figure}[!t]
\includegraphics*[width=80mm]{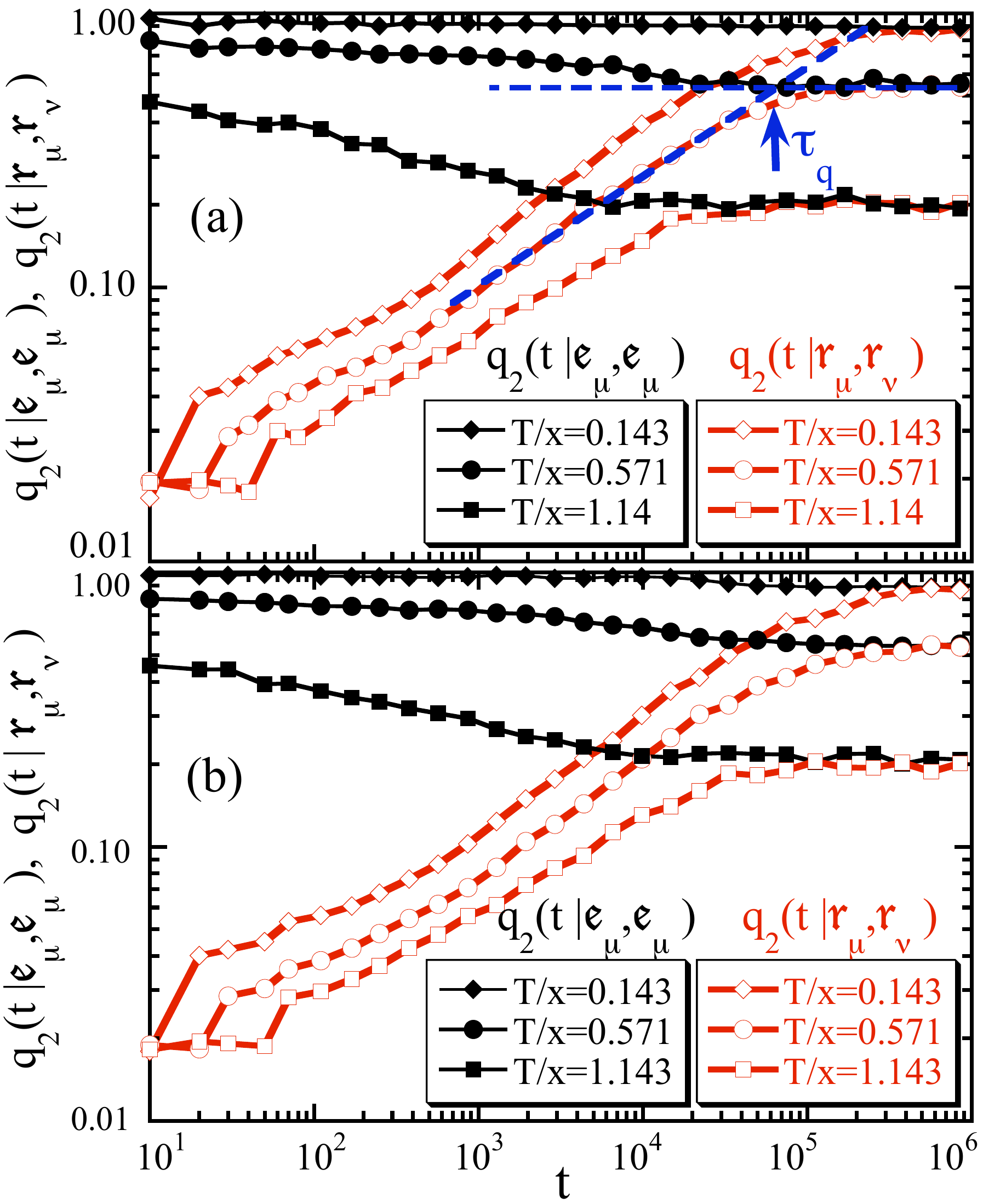}
\caption{(Color online)  (a) Plots of  ${q}_2(t_n\mid \mathfrak{r}_\mu,\mathfrak{r}_\nu)$ and of ${q}_2(t\mid \mathfrak{e}_\mu,\mathfrak{e}_\mu)$ vs $t$, in MC sweeps, for the values of $T/x$ shown,  $x=0.35$, $L=8$, and $H=0.0$. The procedure that was followed to arrive at values for $\tau_q$ is illustrated.
For  ${q}_2(t_n\mid \mathfrak{e}_\mu,\mathfrak{e}_\mu)$, equilibration was allowed to proceed for $5\times 10^5$ MC sweeps before measurements were taken.
Lines are guides to the eye. Error bars are given by the size of the icons. (b) Same as in (a) but for $H=0.2$.}
\label{q2t}
\end{figure}

During equilibration, ${q}_2(t\mid \mathfrak{r}_\mu,\mathfrak{r}_\nu)$ is expected to increase up to its equilibrium value, $q_2$. 
In Fig. \ref{q2t}a, ${q}_2(t\mid \mathfrak{r}_\mu,\mathfrak{r}_\nu)$ is given for $T/x=0.571$ and $T/x=1.14$, at $H=0$. In Fig. \ref{q2t}b,  $H=0.2$, but everything else is as in Fig.  \ref{q2t}a.

Finally,  assume  two replicas start evolving independently from the same  \emph{equilibrium}  state $\mathfrak{e}_\mu$ at time $t=0$. That is, any state $\mathfrak{e}_\mu$ is selected from the sequence of states the system of interest goes through after thermal equilibrium has been reached.
The time dependent equilibrium correlation function ${q}_2(t\mid \mathfrak{e}_\mu,\mathfrak{e}_\mu)$
is the average of $q^2$ at time $t$ over all sample realizations. Again, $ \mathfrak{e}_\mu,\mathfrak{e}_\mu$ appear in ${q}_2(t\mid \mathfrak{e}_\mu,\mathfrak{e}_\mu)$ only to remind us that both replica evolutions start at $t=0$ from the same $\mathfrak{e}_\mu$ equilibrium state.

Note that
${q}_2(0\mid \mathfrak{e}_\mu,\mathfrak{e}_\mu)=1$, and that ergodicity implies ${q}_2(t\mid \mathfrak{e}_\mu,\mathfrak{e}_\mu)\rightarrow q_2$ as $t\rightarrow \infty$.  Therefore, ${q}_2(t\mid \mathfrak{e}_\mu,\mathfrak{e}_\mu)$ is expected to be an upper bound to $q_2$. Plots of \mbox{${q}_2(t\mid \mathfrak{e}_\mu,\mathfrak{e}_\mu)$} are shown in Fig. \ref{q2t}a for $T/x=0.571$ and $T/x=1.14$ at $H=0$. In Fig. \ref{q2t}b,  $H=0.2$, but everything else is as in Fig.  \ref{q2t}a.

 \begin{table}\footnotesize
\caption{Number $\tau_s$ of Monte Carlo sweeps (MCS) allowed, first for equilibration and, subsequently, for averaging over equilibrium, and number $N_s$ of samples for the SK model and for DID models of various linear sizes $L$. For the SK model, $L^3$ is the number of spins. For the DID model, $L$ is given in units of the lattice constant,
each site is occupied with $0.35$ probability, the temperature $T$ fulfills $0.05\leq T\leq 2.0$, and  the temperature spacing between systems in the tempered MC runs is $\Delta T=0.05$.}
\begin{ruledtabular}
\begin{tabular}{|c|c|c|c|}
Model & $L$ & $\tau_s$  & $N_s$   \\
\hline
{SK} & $4$ & $500$ &$10^3$   \\
{SK} & $6$ & $10^3$ &$10^3$  \\
{SK} & $8$ & $5\times 10^3$ &$10^3$  \\
{SK} & $12$ & $ 10^4$ &$500$  \\
\hline
{DID} & $4$ & $10^4$ &$5\times 10^3$  \\
{DID} & $6$ & $10^5$ &$3\times 10^3$  \\
{DID} & $8$ & $10^6$ &$10^3$  \\
{DID} & $10$ & $10^7$ &$300$  \\
\end{tabular}
\end{ruledtabular}
\end{table}

A measure $\tau_q$ of  equilibration times in  tempered MC evolutions, under the conditions specified in Table I, is defined graphically  in Fig. \ref{q2t}a.  
It turns out that $\tau_q\approx  10^2,\; 3\times 10^3,\; 5\times 10^4,\; 10^6$ for $L=4,\; 6,\; 8, \;10$, respectively, for the DID model.  For equilibrium observations below, all MC runs went on for $2\tau_s$ MCS. Values of $\tau_s$ are given in Table I. They fulfill $\tau_s\gg\tau_q$. Equilibrium was achieved in the first half of each run, that is while $t<\tau_s$. All time averages for the calculation of equilibrium values were taken while $\tau_s<t<2\tau_s$.

The following rules for the time evolution of ${q}_2(t\mid \mathfrak{r}_\mu,\mathfrak{r}_\nu)$ under a tempered MC algorithm are noted in passing. The first rule, ${q}_2(0\mid \mathfrak{r}_\mu,\mathfrak{r}_\nu)=1/N$, which follows from the fact that 
spin configurations are initially random, is exact. The second rule,  that  ${q}_2(t\mid \mathfrak{r}_\mu,\mathfrak{r}_\nu)\sim t^{\zeta_L(T)}$ when $10\lesssim t\lesssim \tau_q$, and $\zeta_L(T)\simeq 0.4$ (weakly dependent on $T$ and $L$), follows from plots of ${q}_2(0\mid \mathfrak{r}_\mu,\mathfrak{r}_\nu)$ vs $t$, such as the ones shown in Figs.   \ref{q2t}a and \ref{q2t}b. Further digression into equilibration behavior under \emph{tempered} MC rules is beyond our aim here, which is simply to determine equilibration times.

\begin{figure}[!t]
\includegraphics*[width=80mm]{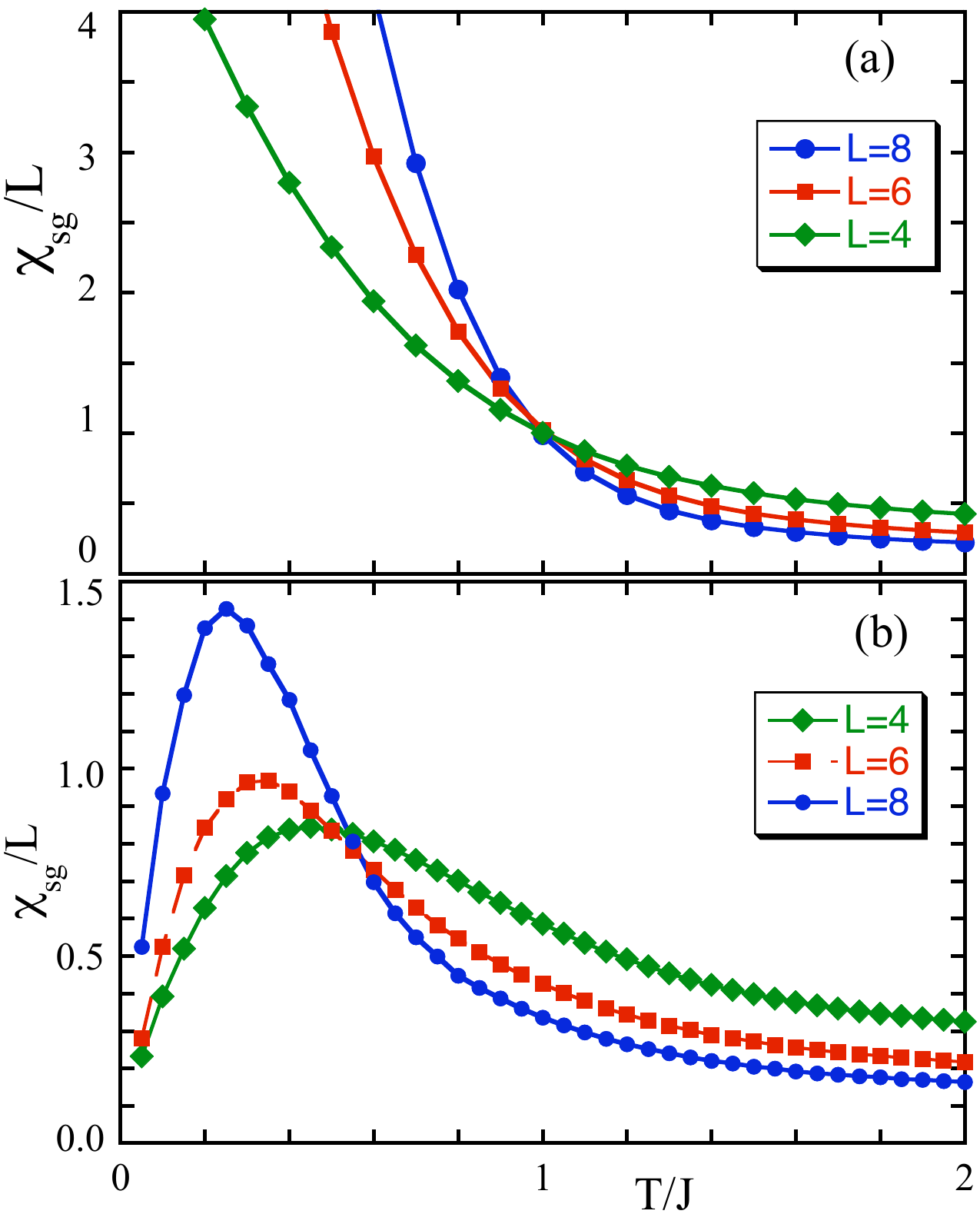}
\caption{(Color online) (a) Plots of $\chi_{sg}/ L$ vs $T/J$ for the SK model at $H=0$ for the values of $L$ shown in the graph. All error bars are much smaller than icon sizes.
(b) Same as in (a) but for $H=0.4J$.}
\label{uno}
\end{figure}

\section{ Equilibrium results}
\label{er}

Equilibrium results obtained from tempered Monte Carlo simulations are reported in this section. These results are for both site-diluted DID models and SK models.
The SK model, in which an AT line is known to exist, is examined for comparison purposes. 

All the data given here for DID models is for $x=0.35$. This is well below $x_c$ ($\simeq 0.65$), in a regime where DID models on SC lattices have been shown\cite{alonso2010} to have an SG phase if $H=0$. Furthermore,\cite{alonso2010} $T_{sg}=1.0(1)x$.

In the search  for the existence of an AT line in DID models, I apply well known criteria.\cite{jorgo} Let 
\begin{equation} 
\chi (\textbf{k})=N^{-1}\sum_{ij} [\langle \delta \phi_i \delta\phi_j\rangle  ]_{av}e^{i\textbf{k}\cdot\textbf{r}_{ij}},
\label{chisg}
\end{equation}
where $\delta \phi_i= \phi_i - \langle \phi_i  \rangle$,
and ${\bf k} =(2\pi/L,0,0)$, perpendicular to all spin directions. Note $\chi(0)$ is the spin-glass susceptibility, $\chi_{sg}$.

In the paramagnetic phase, short range spin-spin correlations imply $\chi_{sg}$ is finite, but $\chi_{sg}\rightarrow\infty $ as the PM-SG critical point is approached. At the critical point,  $\chi_{sg}/L$ remains finite as $L\rightarrow\infty$ in the SK model.\cite{skscale}
Plots of $\chi_{sg}/L$ vs $T$, shown  in Fig. \ref{uno}a for $H=0$ and various values of $L$, exhibit the expected behavior. Similar plots for $H=0.4$ are shown in Fig. \ref{uno}b. 
Clearly, $\chi_{sg}/L$ curves for various values of $L$ do cross, as expected for the SK model, at a nonvanishing value of $T$. Furthermore, they do so at
$T/J=0.55(5)$, which is, within errors, on the AT line.\cite{at,japon}

\begin{figure}[!t]
\includegraphics*[width=80mm]{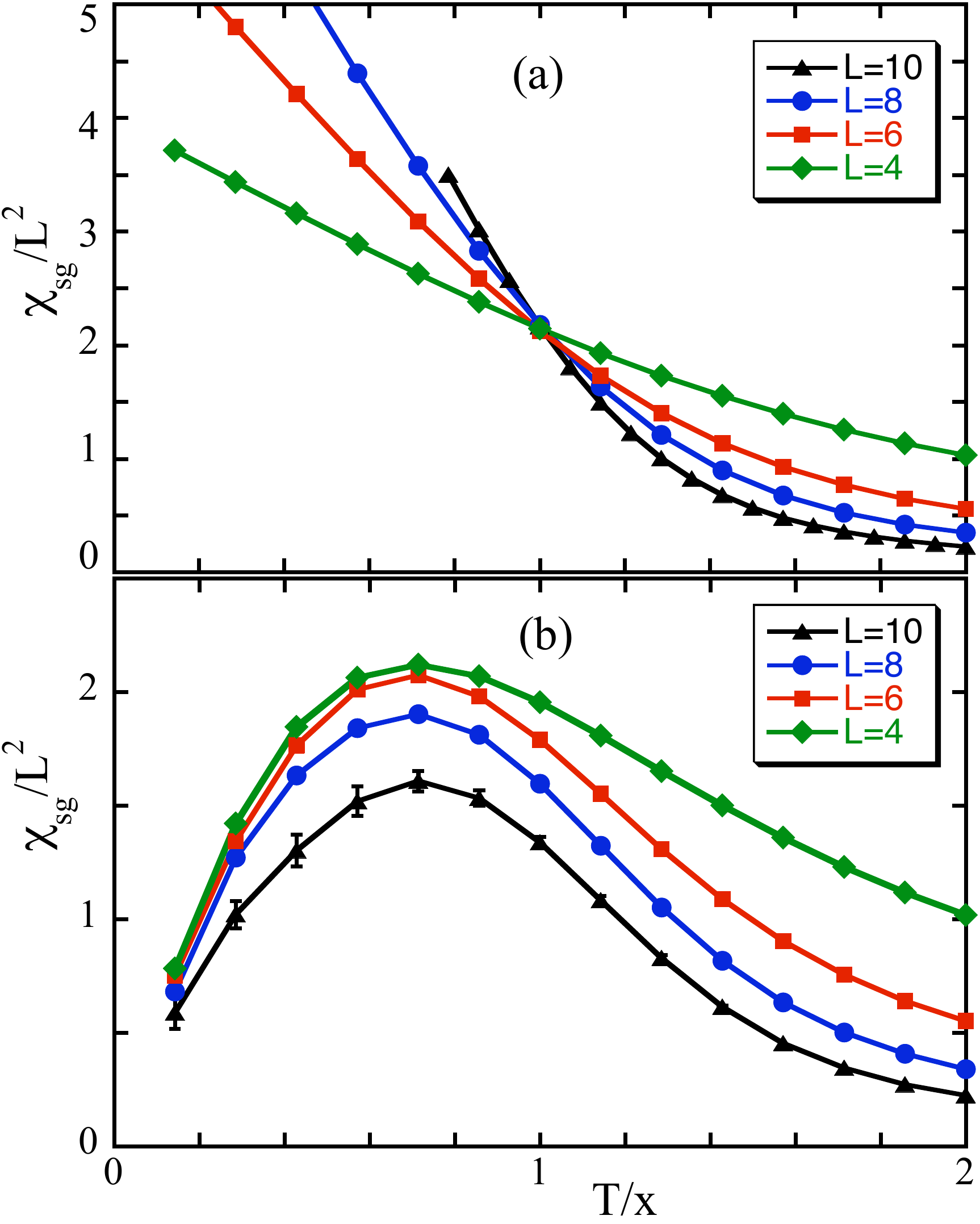}
\caption{(Color online) (a) Plots of $\chi_{sg}/L^2$ vs $T$ for the DID model, for $x=0.35$, $H=0$, and the $L$ values shown in the graph. (b) Same as in (a) but
for $H=0.1$. Error bars show wherever they protrude beyond icons.  }
\label{tres}
\end{figure}

For DID models, one must first decide how to scale $\chi_{sg}$.
Recall that, quite generally, finite size scaling predicts a finite limit of $\chi_{sg}/L^{2-\eta}$ at a critical point as $L\rightarrow\infty$. Furthermore,\cite{alonso2010} $\eta\simeq 0$ in DID models. Plots of $\chi_{sg}/L^{2}$ vs $T$ for $H=0$ and various values of $L$ are seen to cross, as expected, at $T/x\simeq 1.0$ in Fig. \ref{tres}a. 

Not knowing in advance the value of $\eta$ for the hypothetical AT line in DID systems, universality is next assumed. Thus  $\eta =0$ is assumed to hold for $H\neq 0$ as well. 
To probe for an AT line, I vary $T$ with $H>0$ constant. One does not want to miss the AT line by choosing too large a value of $H$. I let $H=0.1$. Since $T_{sg}\simeq x$ for $x\lesssim 0.5$ and $H=0$, and $x=0.35$  has been chosen everywhere, $H=0.1$ gives a Zeeman energy of $0.3k_BT_{sg}$ approximately, which is a rather small field. (For comparison, recall that $H$ along the AT line increases beyond\cite{at} $H=3k_BT_{sg}$ as $T\rightarrow 0$ in the SK model.)

Plots of $\chi_{sg}/L^2$ vs $T$ at $H=0.1$ are shown in Fig. \ref{tres}b.
These results show the AT line, if there is one, is restricted to $H<0.1$, that is, to $\mid H\mid \lesssim 0.3T_{sg}$.   

If instead of $\eta=0$ one uses $\eta=-0.3$, from the table given in Ref. \onlinecite{katz0} for the EA model in 3D, the plots in Figs.  \ref{tres}a and  \ref{tres}b are slightly modified. For $H=0$, curves for different values of $L$ would then cross at $T/x=0.8$, instead of at  $T/x=1.0$, as in Fig. \ref{tres}a.  For $H=0.1$ the main effect is to spread all curves shown in Fig. \ref{tres}b further apart, thus strengthening the conclusion drawn above about the AT line.

 \begin{figure}[!t]
\includegraphics*[width=80mm]{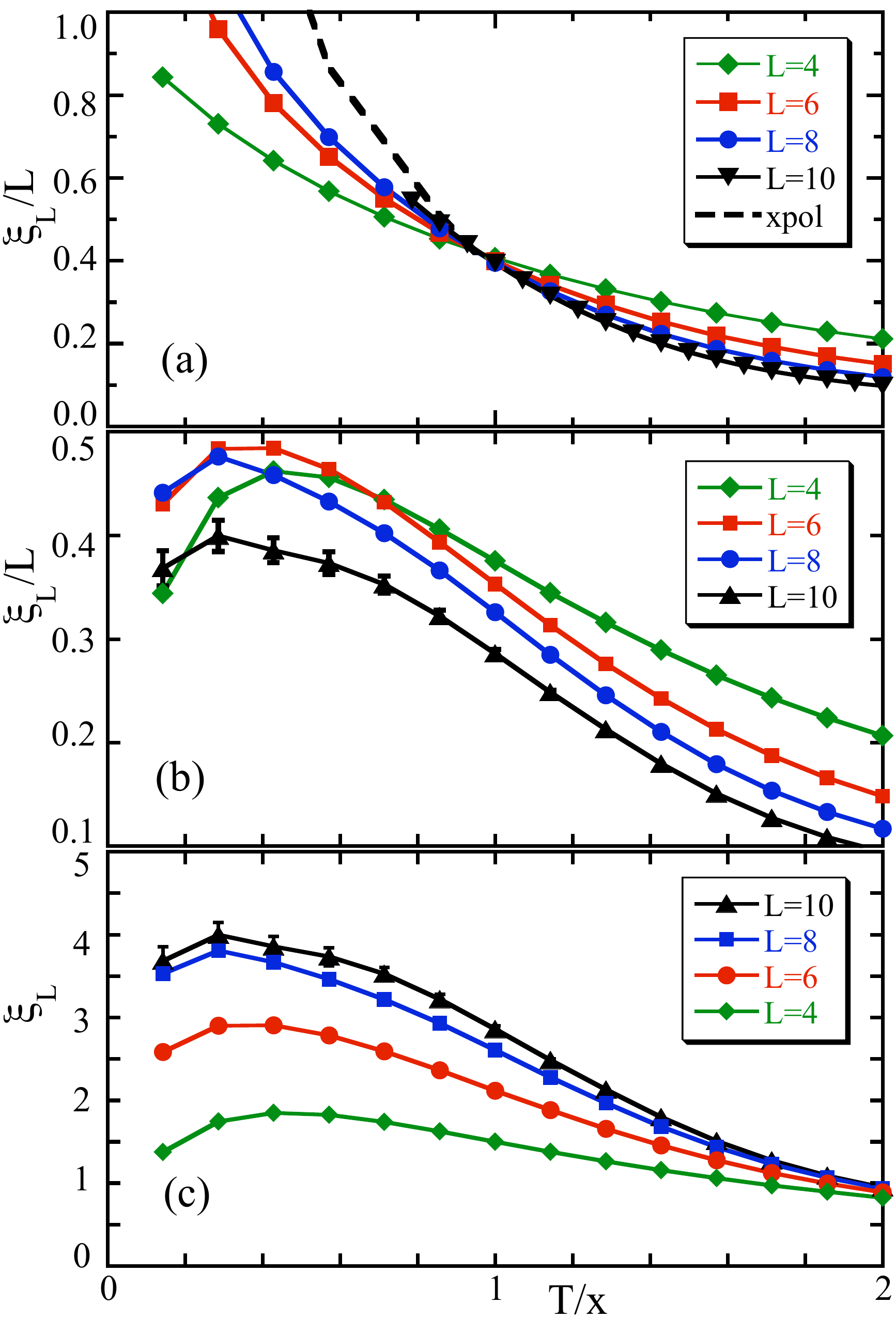}
\caption{(Color online)   (a) Plots of $\xi_L/L$ vs $T$ for the DID model, for $x=0.35$, $H=0$, and the $L$ values shown in the graph. Error bars show wherever they protrude beyond icons. (b) Same as in (a) but
for $H=0.1$.  (c) Same as in (b) but for $\xi_L$ instead of $\xi_L/L$.  }
\label{xiPAD}
\end{figure}

The correlation length $\xi$ is more convenient than $\chi_{sg}$ to work with, because $\xi /L\rightarrow $ remains finite at the critical point as $L\rightarrow \infty$  while $\xi /L\rightarrow 0$ in the paramagnetic phase. Diagnostics with $\xi /L$ is thus free from errors in the value of $\eta$. Let
 \begin{equation}
\overline{\xi}^2_L=  \frac { \sum_{ij}    (\hat{ \textbf {k}}  \cdot  \textbf { r}_{ij})^2 \langle \delta \phi_i \delta \phi_j\rangle }{\sum_{ij} \langle \delta\phi_i \delta \phi_j\rangle },
\label{sqr}
\end{equation}
where $\hat{\textbf{k}}$ is a unit vector along $\textbf{k}$,
and the $L$ subscript is a reminder of the fact that, inevitably, the sum in the equation is performed over \emph{finite} $L^3$ size systems.
Obviously, $\overline{\xi}_L$ is a correlation length measured along the $\textbf{k}$ direction.

Numerical computations of the \emph{double} sum in Eq. (\ref{sqr}) are however time consuming. In addition, $\overline{\xi}_L$ is not well defined if $\langle \delta\phi_i \delta \phi_j\rangle$ decays (as it does\cite{droplet2} in the SG phase) more slowly than $r_{ij}^{-p}$ and \mbox{$p<d+2$}.
Both difficulties are avoided with the definition,\cite{cooper} 
\begin{equation} 
\xi^2_L=\frac {1 } {4 \sin^2  ( k /2)}  { \left[ \frac{ \chi (0)  }{ \mid \chi (\textbf{k})\mid     }  -1 \right] }.
\label{nosa}
\end{equation}
Note that $\xi_L\rightarrow \overline{\xi}_L/\sqrt{2}$ as $\xi_L /L\rightarrow 0$ in the macroscopic limit if $\xi_L$ is finite, since (i) \mbox{$\exp i\textbf{k}\cdot\textbf{r}_{ij}$} can then be replaced by \mbox{$1+  i\textbf{k}\cdot\textbf{r}_{ij} -
(\textbf{k}\cdot\textbf{r}_{ij})^2/2$} in Eq. (\ref{chisg}), and (ii) $(2/k)\sin (k/2)\rightarrow 1$ then.
Thus, Eqs. (\ref{sqr}) and (\ref{nosa}) are qualitatively equal in the paramagnetic phase. Equation (\ref{nosa}) is therefore, as has become customary in SG work,\cite{balle,jorgo,alonso2010} adopted here as the  definition of correlation length. 

In the paramagnetic phase, $\xi_{L}/L \rightarrow 0$ as $L\rightarrow \infty$.
What various assumptions about the SG phase imply for the variation of $\xi_L /L$ with $L$ is discussed in some detail in Sec. VB of Ref. \onlinecite{alonso2010}.
In short, (i) $d_L<3$ (recall $d_L$ is the lower critical dimension) implies $\xi_L/L\rightarrow \infty $ (and a nonvanishing $\chi_{sg}/N$) in the SG phase as $L\rightarrow \infty$, (ii) $d_L=3$ implies $\xi_L/L$ remains finite 
(and $\chi_{sg}/N\rightarrow 0$ but $\chi_{sg}\rightarrow \infty$) in the SG phase as $L\rightarrow \infty$.

Plots of $\xi_L/L$ vs $T$ for the DID model at $H=0$ and $x=0.35$ are shown in Fig. \ref{xiPAD}a. The  $L\rightarrow\infty$ limit of $\xi_L/L$, obtained from $1/L\rightarrow 0$ extrapolations of $\xi_L/L$  in Ref. \onlinecite{alonso2010}, is also shown in Fig. \ref{xiPAD}a. 

Plots of $\xi_L/L$ vs $T$ for the DID model at $H=0.1$, are shown in Fig.  \ref{xiPAD}b. Curves do cross for the smaller values of $L$, but the trend is reversed for larger $L$. 
Then, $\xi_L/L$ decrease as $L$ increases, at least for the temperatures studied. 
With a confidence level above $99\%$, $95\%$, and $85\%$, $\xi_L/L$ is smaller for $L=10$  than for $L=8$, at $T/x\geq 0.43$, $T/x= 0.28$, and, $T/x= 0.14$, respectively.
As for $\chi_{sg}$ above, this is the behavior one expects of $\xi_L/L$ if there is no AT line. 

Plots of $\xi_L$ vs $T$ on Fig.  \ref{xiPAD}c are perhaps more revealing. Clearly, $\xi_L$ saturates for all $T$ to a finite value for $L\gtrsim 8$, as one expects from a paramagnetic phase. 

We end this section with a comment about spatial anisotropy in DID systems. Recall interactions along the $z$-direction, parallel to the spins axis, are twice as large as in a perpendicular direction. The ``longitudinal'' (for $\textbf {k}$ along the $z$-direction) correlation length $\xi^l_L$ is consequently somewhat larger, up to twice as large for high temperatures, than the transverse correlation length $\xi_L$. More importantly, $\xi^l_L/L$ appears to suffer from finite size scaling corrections in a way that $\xi_L/L$ does not:
whereas $\xi_L/L$ for systems of various sizes all cross at approximately the same temperature in Fig. \ref{xiPAD}a, 
$\xi^l_L/L$ do not quite do so for $L=4,\;6,\;8$ and $10$. The crossing points for $\xi^l_L/L$ drift towards
$T_{sg}$ as system sizes increase. For this reason, transverse correlation lengths are more convenient to work with. For $H\geq 0.1$, we find no qualitative difference between $\xi^l_L$ and $\xi_L$.
 
\section{conclusions}
\label{concl}

Spin-glass behavior in an applied magnetic field $H$ has been studied. More specifically, I have numerically probed a site-diluted Ising dipole model of magnetic dipoles for the 
existence of an Almeida-Thouless phase-boundary line. 
This DID model has been previously shown\cite{alonso2010} to have, in three dimensions, at $H=0$ and low temperatures, (i) an AF phase for $x>x_c$, where $x_c=0.65(5)$, (ii) a (marginal) SG phase for $x<x_c$, that is $d_L\simeq 3$, (iii) a behavior for $x\ll x_c$ that is independent of lattice structure and depends on $x$ and $T$ only through $T/x$, and (iv) $T_{sg}/x\simeq 1$.  
Spin-glass behavior as a function of $T$ and $H$ can therefore be inferred for all $x\ll x_c$ from that at a single small value of $x$. 

Here, equilibrium results, from tempered Monte Carlo simulations, are reported for $\chi_{sg}/L^{2-\eta}$ and  $\xi_L/L$ for the DID model at $x=0.35$, various temperatures and system sizes, at $H=0$ and $H=0.1$.
The criterion for the existence of an AT line,  that $\chi_{sg}/L^{2-\eta}$ and  $\xi_L/L$ be independent of $L$ at the critical point, has been shown here to work well for (i) the SK model at $H=0$ and $H=0.4J$, that is, $H=0.4T_{sg}$, for which the answer has long been known,\cite{at} as well as (ii) for the DID model at $H=0$. For $H=0.1$, that is, $H\simeq 0.3T_{sg}$, the trend observed in the data is clearly  away from $\chi_{sg}/L^{2-\eta}$ or  $\xi_L/L$ becoming independent of $L$ at any $T$ as $L\rightarrow \infty$. Indeed,  $\xi_L$ saturates to a finite value beyond $L\simeq 8$ for all $T$. 
All of this is consistent with the absence of an AT phase boundary line in the DID model, at least above $H\gtrsim 0.3T_c$.

\acknowledgments
I am grateful to J. J. Alonso and to F. Luis for helpful remarks. This study was funded by Grant FIS2009-08451, from the Ministerio de Ciencia e Innovaci\'on of Spain.


\begin{thebibliography}{99}
\bibitem{earlydl}There is a large spread in the predicted values of $d_L$ for the EA model, from values as large as  $d_L\simeq 4$, in Ref. \onlinecite {Fisch},  to values as low as $d_L= 2$ for $J=\pm 1$, in Refs. \onlinecite{kardar}
\bibitem{stiff2} This expression is a simplified version of the one given in Ref. [\onlinecite{stiff}].
\bibitem{Hdef}The units of $H$ are given in Sec. \ref{mo}.
\end{thebibliography}
\end{document}